\documentclass[aps,prl,nofootinbib,amsmath,twocolumn,preprintnumbers]{revtex4-1}
\usepackage{amssymb,esvect,amsmath,graphicx,latexsym,amsthm,slashed,eso-pic,hyperref}
\usepackage[hang,flushmargin]{footmisc}
\usepackage[final]{pdfpages}
  \DeclareMathOperator{\mev}{MeV} \DeclareMathOperator{\gev}{GeV} \DeclareMathOperator{\tev}{TeV}     \DeclareMathOperator{\fb}{fb} \DeclareMathOperator{\pb}{pb}     \DeclareMathOperator{\few}{few}  \DeclareMathOperator{\ifb}{fb^{-1}} 
      \newcommand{\cL}{{\cal L}}   \newcommand{\cO}{{\cal O}}    
\newcommand{\ep}{\epsilon}

\newcommand{\pL}{\left(} \newcommand{\pR}{\right)} \newcommand{\bL}{\left[} \newcommand{\bR}{\right]} \newcommand{\cbL}{\left\{}  \newcommand{\mL}{\left|} \newcommand{\mR}{\right|}
\newcommand{\beq}{\begin{equation}} \newcommand{\eeq}{\end{equation}}
\newcommand{\bea}{\begin{eqnarray}} \newcommand{\eea}{\end{eqnarray}}
\newcommand{\alg}[1]{\begin{align} \begin{split} #1 \end{split}  \end{align}}

\newcommand{\Eq}[1]{Eq.~(\ref{#1})}  
  
\newcommand{\Fig}[1]{Fig.~\ref{#1}} 
\newcommand{\Tab}[1]{Tab.~\ref{#1}}

\DeclareMathOperator{\br}{Br}

\begin{document}
\preprint{YITP-SB-15-47}

\title{Singlet Scalar Resonances and the Diphoton Excess}
\author{Samuel D.~McDermott, Patrick Meade, and Harikrishnan Ramani}
\affiliation{C.~N.~Yang Institute for Theoretical Physics, Stony Brook, NY 11794}
\date{\today}

\begin{abstract}
ATLAS and CMS recently released the first results of searches for diphoton resonances in 13 TeV data, revealing a modest excess at an invariant mass of approximately 750 GeV. We find that it is generically possible that a singlet scalar resonance is the origin of the excess while avoiding all other constraints. We highlight some of the implications of this model and how compatible it is with certain features of the experimental results. In particular, we find that the very large total width of the excess is difficult to explain with loop-level decays alone, pointing to other interesting bounds and signals if this feature of the data persists. Finally we comment on the robust $Z\gamma$ signature that will always accompany the model we investigate.
\end{abstract}
\maketitle

\section{Introduction}
\label{s.intro} \setcounter{equation}{0}

The excess found in diphoton final states in the $13\tev$ ATLAS and CMS data at 750 GeV~\cite{atlasres,cmsres} presents an interesting model-building challenge\footnote{While this paper was in preparation a number of attempts at explaining the diphoton excess appeared~\cite{everyone}}.  ATLAS and CMS both characterize the events that make up their signal to have the same composition as the background in sidebands~\cite{atlasres,cmsres}.  Therefore, we assume that this peak in the event spectrum comes from the direct production of a resonance rather than a cascade decay.

If the signal is caused by a resonance that decays to diphotons, the Landau-Yang theorem~\cite{landauyang} restricts the spin of the resonance to 0 or 2.  The spin 2 possibility was preliminarily investigated by CMS~\cite{cmsres}, and, though interesting, it is difficult to construct a model that satisfies all constraints.  However, an example of a scalar resonance which decays to diphotons is already provided by the SM Higgs, illustrating that it is straightforward to construct a  ``cousin" of the Higgs to explain this excess.  

Models of a scalar resonance which explain the excess can come from sectors with a wide variety of field content and quantum number assignments~\cite{everyone}.  The simplest possibility which avoids many correlated bounds is a resonance that is a singlet under the Standard Model gauge group.  This implies that the coupling to protons and photons is generated by loops of new non-Standard Model particles that are colored and charged. 

As discussed in more detail below, the width of the excess preferred by ATLAS and CMS~\cite{atlasres,cmsres} immediately implies additional constraints on singlet models. The preferred fit from ATLAS has a width of 45 GeV~\cite{atlasres} with a local significance of 3.9 $\sigma$, although this represents only a marginal improvement over a narrow-width model.  CMS slightly prefers a narrow width~\cite{cmsres}, but overall has a smaller number of events, which can be partially attributed to their lower luminosity. The model point favored by CMS data has a width of $\mathcal{O}(100\mev)$, with an excess of 2.6 $\sigma$.  However, CMS is compatible at a similar level of confidence with a width of 42 GeV.  It should be noted that ATLAS, while compatible with narrow width, prefers a larger width for several reasons.  In the narrow width model, ATLAS finds a pull based on marginalizing over the width. This indicates a resonance width larger than the experimental resolution of $5.3\gev$~\cite{atlasres}.  Additionally, when comparing the excess between 13 TeV and 8 TeV, ATLAS find that the narrow width is only compatible at the 2.2$\sigma$ level whereas the larger width is compatible with a smaller 1.4$\sigma$ tension.  Given the limited data it is therefore still possible to have a narrow width (indicative, as we discuss below, of strictly loop-induced processes), but it is more experimentally favored to have a larger width.   

This brings singlet models under some tension. The number of observed photons is given by
\begin{equation}
N_{\gamma\gamma}=\sigma_{\rm prod} \times \frac{\Gamma_{\gamma\gamma}}{\Gamma_{\rm tot}} \times \mathcal{L}\times \epsilon \times A,
\end{equation}
where $\sigma_{\rm prod}$ is the production cross section for $pp\rightarrow\,\mathrm{resonance}$, $\Gamma_{\gamma\gamma}$ is the partial width for the resonance to decay into photons, $\Gamma_{\rm tot}$ is the full width of the resonance, and $ \mathcal{L}\times \epsilon \times A$ are the usual Luminosity/efficiency/acceptance factors dictated by the experiment. We find a rough estimate of the $\epsilon \times A\sim 0.8$ for the ATLAS study~\cite{atlasres} based on a MadGraph~\cite{Alwall:2014hca} parton level simulation. The characteristic size of $\Gamma_{\gamma\gamma}$ coming from a loop induced decay is 
\beq
\Gamma_{\gamma\gamma} \sim \frac{\alpha_{\rm EM}^2 m_S^3}{256\pi^3 m_L^2}
\eeq
where $m_S$ is the mass of the resonance and $m_L$ is the mass of the charged particle in the loop responsible for coupling to photons.  Assuming that no particles that couple the resonance to the SM via a loop provide tree-level decay modes for the resonance, and taking $\mathcal{O}(1)$ couplings, this provides a rough bound on the partial width into diphotons of $\Gamma_{\gamma\gamma}\sim\mathcal{O}(50)$ MeV.  Therefore if $\Gamma_{\rm tot}$ is near the 45 GeV value preferred by ATLAS, in any model with a singlet scalar, the number of diphoton events is suppressed by $\frac{\Gamma_{\gamma\gamma}}{\Gamma_{\rm tot}}\lesssim \mathcal{O}(10^{-3})$.  While this suppression is less severe than it would be for a copy of the Higgs of a similar mass, we show below that generating a large enough total event rate and cross section nevertheless provides some interesting tension even at large coupling.  This provides opportunities for ATLAS and CMS to test this hypothesis in current data.  This also provides bounds on the types of decay modes the resonance can have based on the earlier runs of the LHC.

\section{The ``QL" Model and the Excess}

\begin{figure}[tb]
\begin{center}
\includegraphics{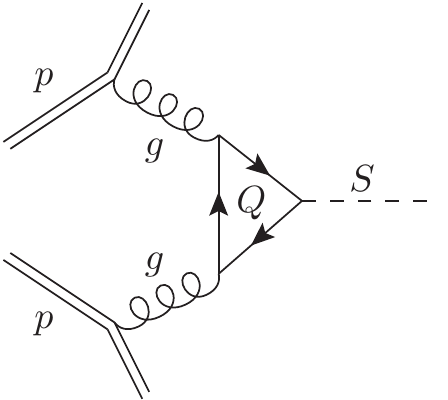}
\caption{Production of $S$ particles from a loop of $Q$'s.}
\label{fig:prod}
\end{center}
\end{figure}

\begin{figure}[tb]
\begin{center}
\includegraphics{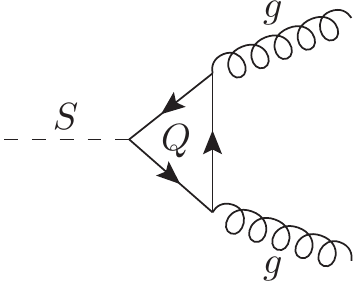}
\includegraphics{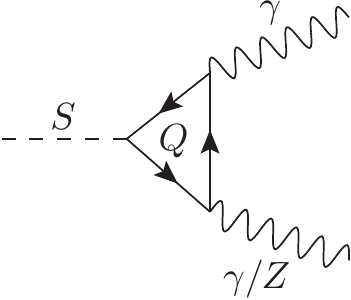}
\includegraphics{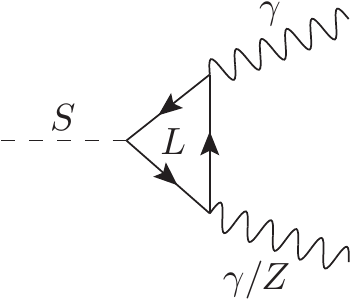}
\caption{Decay of $S$ particles from loops of $Q$'s and $L$'s.}
\label{fig:decay}
\end{center}
\end{figure}

Assuming a singlet scalar $S$, we need to construct a model which couples the $S$ to new non Standard Model particles, allows for production in $pp$ collisions, and leads to diphoton decays. The simplest way to achieve this is to couple $S$ to a vector-like pair of fermions. The most economical model consists of adding a single pair of colored and hypercharged fermions, thus providing for loop-level couplings to gluons and photons. The ratio of these couplings will depend on the charges and masses of the loop particles.  However, to provide for maximum freedom in separately adjusting the partial width of the resonance into gluons and photons, a more universal ``module'' will consist of a colored fermion pair and an uncolored but hypercharged pair.  In principle a colored fermion pair could have hypercharge zero, but this leads to novel collider signatures~\cite{Meade:2011du} which are beyond the scope of this paper.  Therefore we will look at a model where we introduce a vector like fermion $Q$ with SM quantum numbers $({\bf3,1})_{q_Q}$ and another called $L$ transforming as $({\bf1,1})_{q_L}$.  This allows us to both dial the ratio of gluon and photon decays and later easily introduce decay modes for $Q$ and $L$. The Lagrangian for $S$, $Q$, and $L$ (excluding the decays of $Q$ and $L$ and their gauge interactions) is given by
\beq \label{QQS}
\cL_{QL} \supset \frac12 m_S^2 S^2+ y_Q \bar Q Q S +m_Q \bar Q Q +y_L \bar L L S + m_L\bar L L,
\eeq
where we assume that $m_Q,m_L \geq m_S/2$. All $S$ particles are produced as in \Fig{fig:prod}, and decay as in \Fig{fig:decay}.

\begin{figure}[tb]
\begin{center}
\includegraphics[width=.48\textwidth]{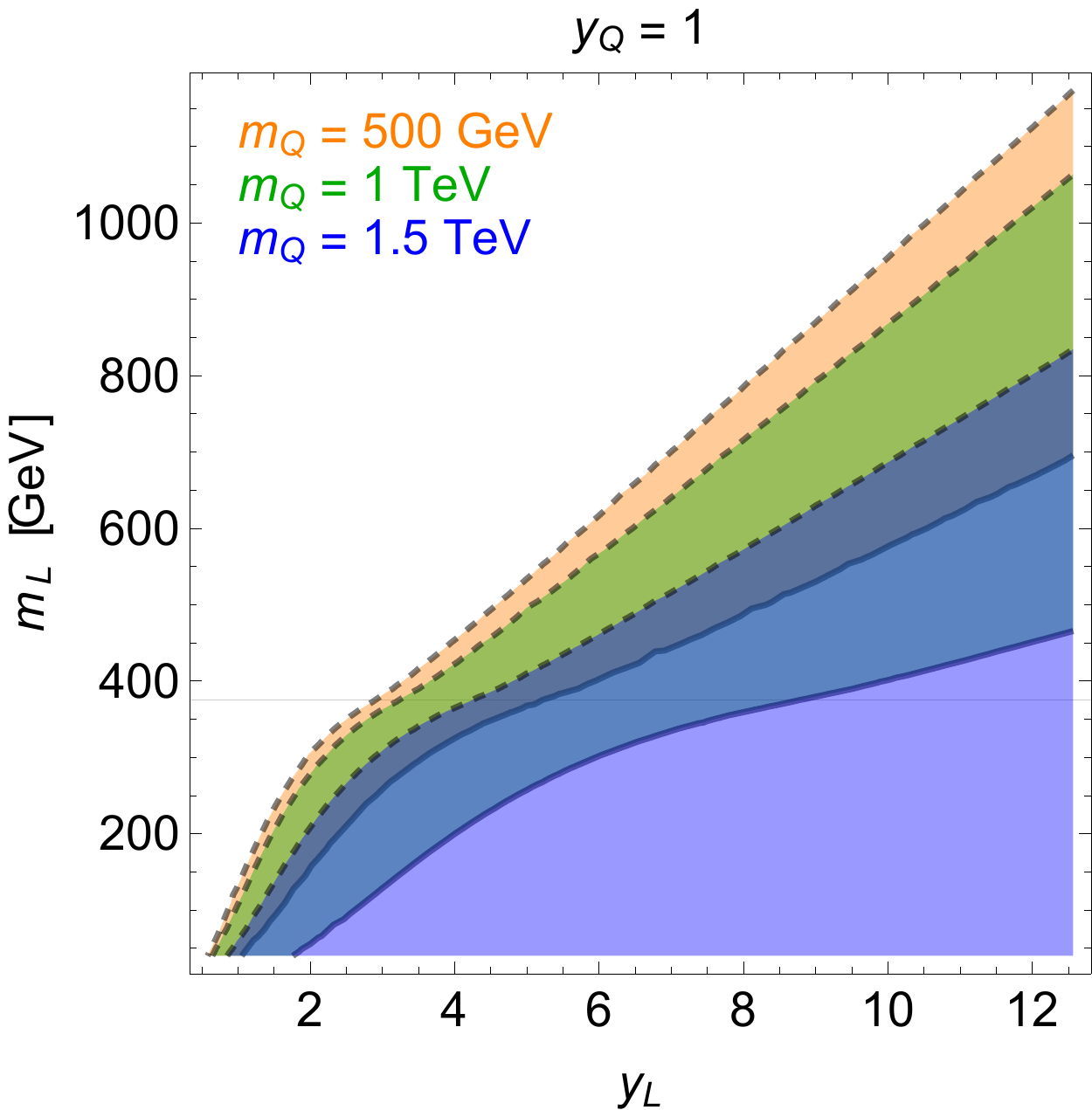}
\caption{Contours of $N_{\gamma \gamma}$ in the $y_L-m_L$ plane for the model in \Eq{eq:mQmLModel} with $q_L=1$. On the dashed (solid) lines we get 5 (15) events with $\cL = 3.2\ifb$. The colored particle mass is fixed by the color coding ($m_Q=500\gev$ is probably ruled out by direct searches, but is included for illustration). Below the faint solid line the $S$ has on-shell decays to $L$.}
\label{fig:mLyLModel}
\end{center}
\end{figure}

We will assume that $m_Q>m_L$, so that the decay to gluons (photons) is effectively mediated strictly by $Q$ ($L$) particles, as in the top left and bottom panels of \Fig{fig:decay}. Therefore the diphoton branching ratio is~\cite{Djouadi:2005gi}
\alg{
\br_{\gamma \gamma}^{QL} &= \frac{\Gamma(S \underset L\to \gamma \gamma)}{\Gamma(S \underset L\to \gamma \gamma)+\Gamma(S \underset Q\to gg)} 
\\& = \frac{ q_L^4 \alpha_{\rm EM}^2 \mL A^L_{1/2}\mR^2 \ep_L^2}{q_L^4 \alpha_{\rm EM}^2 \mL A^L_{1/2}\mR^2  \ep_L^2+2K\alpha_s^2 \mL A^Q_{1/2}\mR^2 \ep_Q^2},
}
where  ``$\underset i\to$'' means ``via a loop of $i$ particles;'' $\ep_i \equiv y_i m_t/y_tm_i $; the ratio $\ep_Q^2/\ep_L^2$ is free but by assumption is less than 1; $A_{1/2}^i$ is a loop function~\cite{Djouadi:2005gi}\footnote{We define $A_{1/2}^i = \frac2{\tau_i^2} \bL \tau_i+(\tau_i-1) f(\tau_i) \bR,$ with $\tau_i = \frac{m_S^2}{4m_i^2}$ and $f(\tau)=\cbL \begin{array}{ll}\arcsin^2(\sqrt{\tau})&\tau\leq 1\\ -\frac14\bL \ln \pL \frac{1 + \sqrt{1-\tau}}{1 - \sqrt{1-\tau}} \pR -i\pi \bR^2 &\tau> 1 \end{array} \right.$ \cite{Djouadi:2005gi}.}; $K$ is the $k$-factor for the two gluon decay; and $\alpha_{\rm EM}, \alpha_s$ are the electromagnetic and strong fine structure constants. The number of diphoton events is
\alg{ \label{eq:mQmLModel}
N_{\gamma \gamma}^{QL} \simeq \frac{\ep_{\rm eff} \cL \bar \sigma \ep_Q^2 }{ 1 + \frac{460}{q_L^4} \frac{\ep_Q^2}{\ep_L^2} \frac{\mL A^Q_{1/2}\mR^2}{\mL A^L_{1/2}\mR^2} } \frac{\mL A^Q_{1/2}\mR^2}{\mL A^t_{1/2}\mR^2},
}
where we take $K=1.7$, $\alpha_{\rm EM} \simeq 1/127$, and  $\alpha_s \simeq 0.92$. We point out that this is essentially independent of the parameters of the $Q$ particle if $\ep_L^2 \ll 460 \ep_Q^2$ (e.g., at large $m_L$ or low $y_L$), since the $\ep_Q$ factors cancel out. However, as the $\br_{\gamma \gamma}$ increases (e.g., with increasing $m_Q$) it eventually asymptotes to 1, and inevitably the decrease in total rate of $S$ production can no longer be accommodated by increasing $\br_{\gamma \gamma}$. In \Fig{fig:mLyLModel} we show the parameter space for this model, with $y_Q=1$. We see that for generic couplings, it is very easy to obtain the correct number of diphoton events. For $m_Q\lesssim 500\gev$ (which is bounded by direct searches for colored states), there is a limited sensitivity to $m_Q$. However, in the range $m_Q \gtrsim 1.5\tev$ the diphoton branching fraction saturates at 1 while the total production of $S$'s is decreased, eventually leading to a loss of signal.

\begin{figure}[bt]
\begin{center}
\includegraphics[width=.48\textwidth]{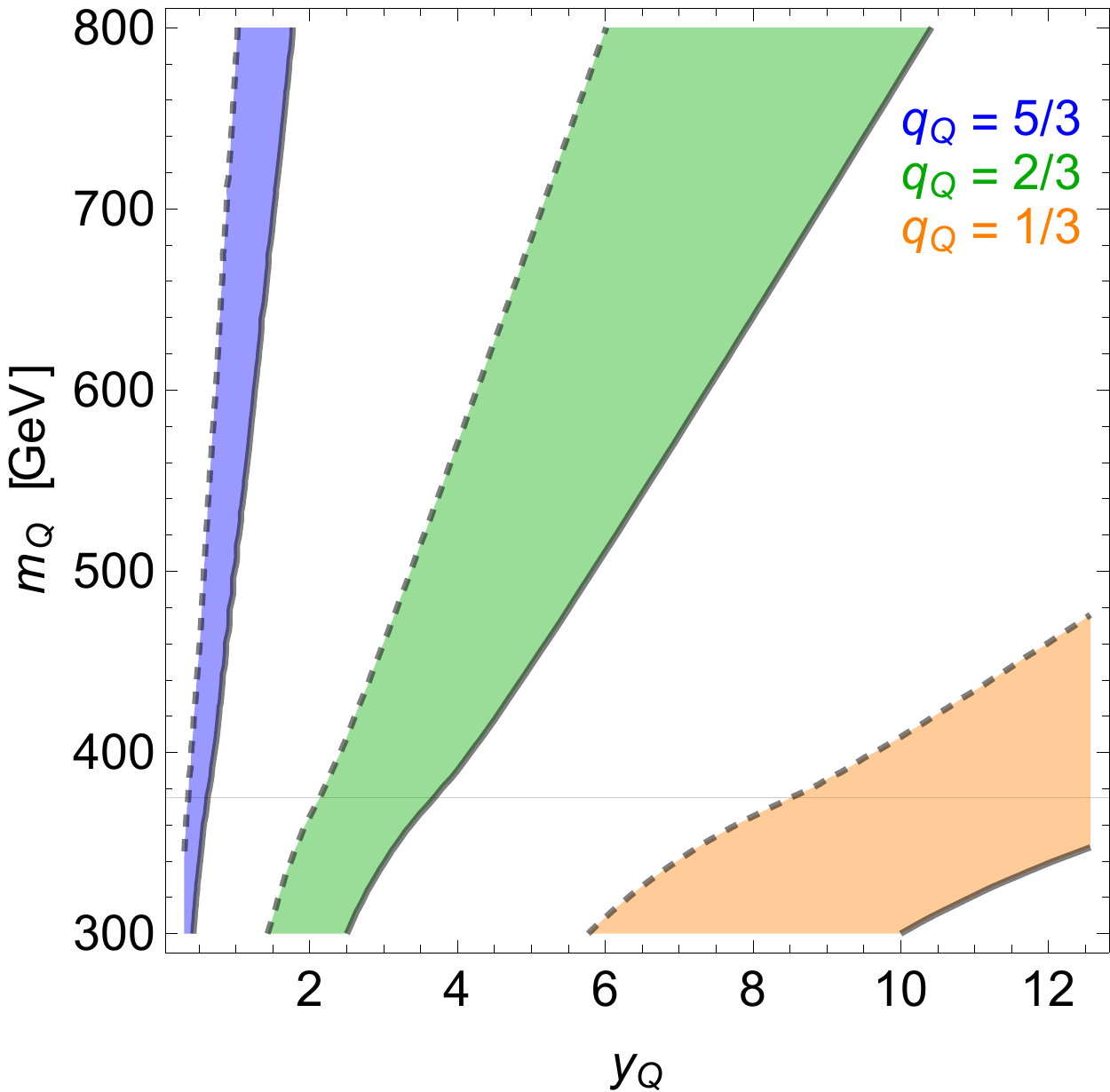}
\caption{Contours of $N_{\gamma \gamma}$ in the $y_Q-m_Q$ plane for the model in \Eq{eq:mQyQModel}. On the dashed (solid) lines we get 5 (15) events with $\cL = 3.2\ifb$. The quark electric charge is fixed by the color coding. There are no additional free parameters. Below the faint solid line, the $S$ has on-shell decays to $Q$.}
\label{fig:mQyQModel}
\end{center}
\end{figure}

As an interesting special example of the ``$QL$ Model,'' we can decouple the $L$ by sending $m_L \to \infty$. This has fewer free parameters, and is correspondingly more minimal than the preceding model. The $S$ will decay through $Q$ loops to pairs of gluons and also pairs of photons, as in the top row of \Fig{fig:decay}, since $q_Q \neq 0$. We find
\beq
\br_{\gamma \gamma}^Q \simeq \frac{9 q_Q^4 \alpha_{\rm EM}^2}{2K\alpha_s^2} \simeq \frac1{260} \pL\frac{q_Q}{2/3} \pR^4,
\eeq
where we assume dijet events dominate. In this case, the number of diphoton events is approximately
\beq \label{eq:mQyQModel}
N_{\gamma \gamma}^Q \simeq \frac{\ep_{\rm eff} \cL \bar \sigma }{ 260 }\pL \frac{y_Qm_t}{y_tm_Q}\pR^2 \pL\frac{q_Q}{2/3} \pR^4 \frac{\mL A^Q_{1/2}\mR^2}{\mL A^t_{1/2}\mR^2}.
\eeq
We show the parameter space for this model in \Fig{fig:mQyQModel} for a variety of choices of $q_Q$. Even in this very simple model, we are able to accommodate the observations at reasonable values of the couplings. This comes at the cost of a large, fixed number of dijet events which may only be borderline compatible with dijet searches. We address these searches in more detail below.

To a very good approximation in the model of \Eq{QQS}, the total width is the sum of the width to gluons via $Q$ loops plus the width to diphoton and photon-Z via $Q$ and $L$ loops. We find
\alg{ \label{char-wid}
\Gamma_S &\simeq \Gamma(S \underset Q\to gg) + \Gamma(S \underset Q\to \gamma \gamma/Z)  + \Gamma(S \underset L\to \gamma \gamma/Z)
\\&\simeq 25 \mev \,  \ep_Q^2 \mL A_{1/2}(\tau_Q) \mR^2 \times
\\&\qquad \times \bL 1  + \pL q_Q^4  +q_L^4 \frac{ \ep_L^2 \mL A_{1/2}(\tau_L) \mR^2}{\ep_Q^2 \mL A_{1/2}(\tau_Q) \mR^2} \pR \middle/290 \bR,
}
where we use $\br_{Z\gamma}=2t^2_W\br_{\gamma \gamma}$ with $t^2_W$ the tangent of the Weinberg angle. Because $|A_{1/2}|\leq2$ when on-shell decays are forbidden and $\ep_i \lesssim \cO(1)$, we see that it is highly nontrivial for $S$ to reproduce the observed width of $\Gamma_S \gg \cO(\few\gev)$ via loop-level decays alone.

\section{Implications of a Broad Width}
As we have seen, the ``QL'' model can account for the number of diphoton events observed in the excesses of ATLAS and CMS.  However, the characteristic total width given in \Eq{char-wid} is far too small to account for the full width if ATLAS's preliminary indications persist.  It is useful therefore to augment the ``QL'' model with some additional contribution to the partial width. We can then find bounds under the assumption that the total width is fixed as an experimental input,
\beq
\Gamma_{\rm tot}=\Gamma_{\gamma\gamma}+\Gamma_{\gamma Z}+\Gamma_{gg}+\Gamma_{X}
\eeq
where $X$ denotes some unknown final state and $\Gamma_{X}$ is as large as necessary to get $\Gamma_{\rm tot}$ to match observations. It is trivial to generate an additional large partial width by introducing another particle that couples to $S$ which allows for tree level decays.  For now we will be agnostic about what this is and simply investigate the constraints on the ``$QL$" sector by increasing the total width of $S$.

\begin{figure}[htbp]
\begin{center}
\includegraphics[width=.48\textwidth]{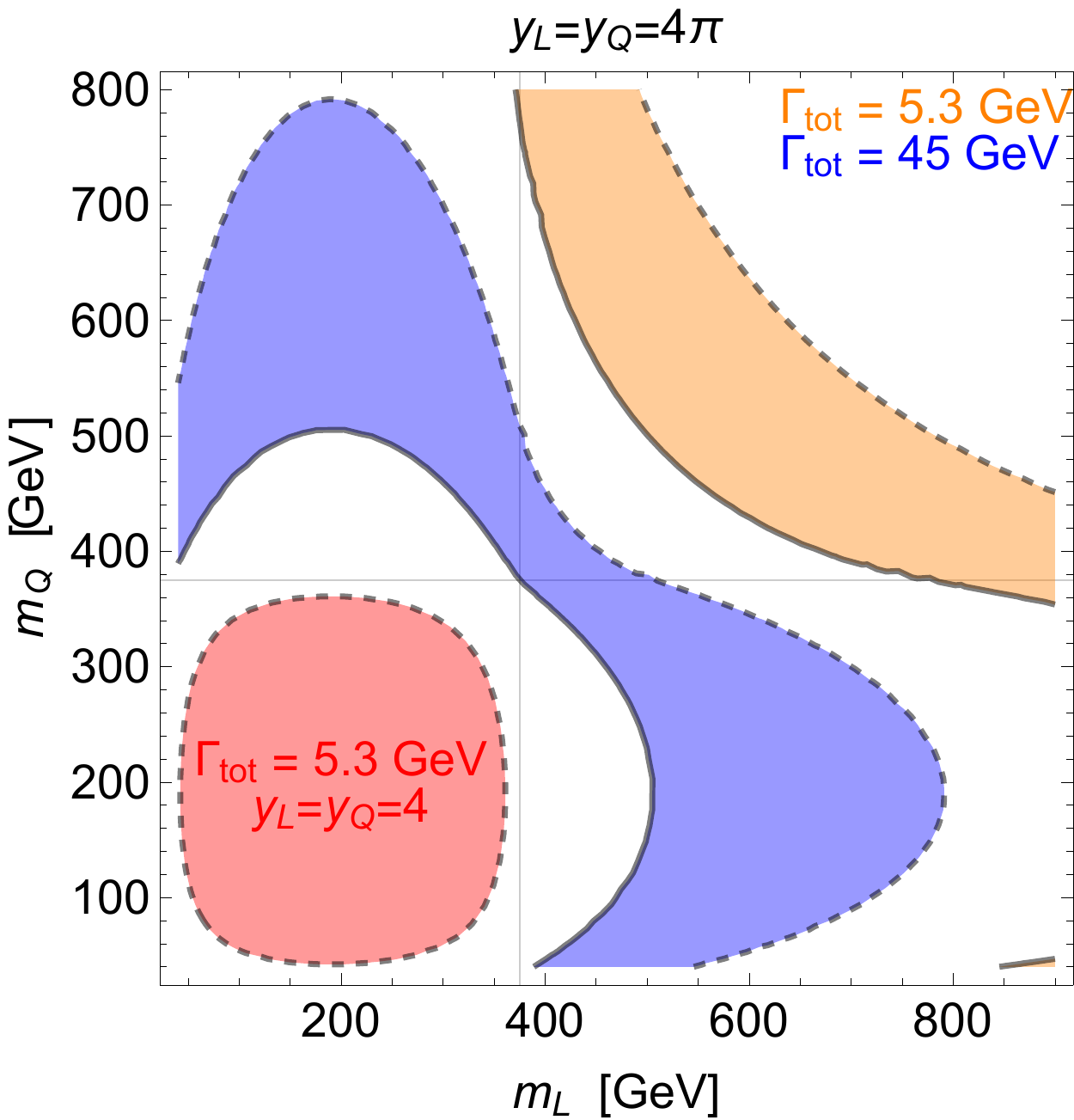}
\caption{Contours of $N_{\gamma \gamma}$ in the $m_L-m_Q$ plane for a fixed width. On the dashed (solid) lines, we get 5 (15) events with a luminosity of $3.2\ifb$, assuming the total width to be fixed to the amount suggested by the color coding. Below and to the left of the faint solid lines, the $S$ has on-shell decays to the new fermions. The experimental resolution at ATLAS is approximately $5\gev$ \cite{atlasres}.}
\label{fig:mQmLFixed}
\end{center}
\end{figure}

For $\Gamma_{\rm tot}$ we take two possibilities, $\Gamma_{\rm tot}=5.3\gev$ and $\Gamma_{\rm tot}=45\gev$, which are the experimental resolution $\sigma_E^{\gamma \gamma}$ and the preferred value of the width from~\cite{atlasres}, respectively. As stated above, ATLAS has a preference for $\Gamma_{\rm tot} \gg  \sigma_E^{\gamma \gamma}$. In \Fig{fig:mQmLFixed} we plot the range of parameters that results in the $S$ resonance giving between 5 to 15 diphotons in the limit of strong coupling between the $S$ and the new fermions. 
As can be seen from \Fig{fig:mQmLFixed}, even in the limit of strong coupling there is essentially no parameter space in this model that can successfully allow for a width of 45 GeV and the right number of diphotons.  Reducing the width to the experimental resolution allows for some additional parameter space, but note that this still requires strong coupling and fixes the mass of the $QL$ both to be less than $800$ GeV to account for the photons seen by ATLAS. This should allow for copious production of $S$ particles at 13 TeV.   This is to be contrasted with what was shown in the previous section where the colored particle mass could be well above a TeV and avoid other potential bounds.  Additionally, the strong coupling limit can not be reduced very much away from $4\pi$.  To demonstrate this, in Figure~\ref{fig:mQmLFixed} we show that if we take the very optimistic width of 5.3 GeV and reduce the couplings only by a factor of $\pi$ such that $y_Q=y_L=4$, there is no viable parameter space remaining. 

If one postulates that the singlet scalar can be strongly coupled to fit the observed width, there will potentially be strong constraints depending on what final state $X$ contributes to $\Gamma_{X}$.  A simple model with a collider stable invisible fermion $\chi$ and a coupling $y_\chi S \bar{\chi}\chi$ would imply that  $\Gamma_{X}$ is an invisible width for $S$.  However, these are already potentially constrained by mono-jet searches at 8 TeV.  In~\cite{Khachatryan:2014rra}, monojet events were analyzed in signal regions for events above different $\slashed E_T$ cuts starting at $250$ GeV.  To translate this into a bound on $\sigma (pp\rightarrow S\rightarrow\mathrm{invisible})$ we perform the following analysis.  We calculate an efficiency which gives the experimental acceptance for each signal region, and normalize against the production cross section for an $S$ without an additional ISR jet:
\begin{equation}
 \epsilon[p_T]=\bL \frac{\sigma(gg\rightarrow S j)}{\sigma(gg\rightarrow S)}\bR(p_T(j)>p_T).
\end{equation}
We assume that $\slashed E_T=p_T(j)$, valid for large transverse momenta. Unfolding with respect to $\epsilon$ gives bounds  at the 95\% C.L. on $\sigma(pp\rightarrow S\rightarrow \mathrm{invisible})=\sigma(pp\rightarrow S)BF(S\rightarrow\mathrm{invisible})$. We provide these bounds for each potential signal region in \Tab{monoX}. 
\begin{table}
    \begin{tabular}{|c|c|}
    \hline
    $\slashed E_T$ threshold [GeV]& $\sigma_{\text{invisible}}$[pb]  \\ \hline
    250 & 3   \\ \hline
    300 & 1.78    \\ \hline
    350 & 0.75   \\ \hline
    400 & 0.65  \\ \hline
    450 & 0.52 \\ \hline
    \end{tabular}
    \caption{Bounds from searches for mono-jet plus $\slashed E_T$ in $8\tev$ data~\cite{Khachatryan:2014rra}.}
    \label{monoX}
\end{table}

Given that the typical decay width for visible gluon and photon decay channels is less than a GeV but $\Gamma_{\gamma\gamma}\sim \cO(5\mev)$, accounting for a 45 GeV total width gives the requirement $\sigma (pp\rightarrow S\rightarrow\mathrm{invisible}) \sim 9\pb$.  Because the weakest possible bound on this model from~\cite{Khachatryan:2014rra} is 3 pb, the possibility for having the rest of the decay width be invisible is ruled out.  If instead the width is assumed to be the experimental resolution, we require $\sigma (pp\rightarrow S\rightarrow\mathrm{invisible}) > 1$ pb. This is compatible with the weaker bins of the analysis, but is ruled out by the higher MET regions.  Therefore we conclude {\it that it is not viable to explain the large width by including only an invisible width for the $S$}: somehow an additional $\mathcal{O}(1)$ branching fraction must go into a visible final state that avoids all other direct bounds.  Regardless, accounting for the larger width requires a much larger superstructure to be compatible with experimental constraints. The model must also be in a strong coupling region with light new colored and charged states.

\section{Other Experimental Constraints}

Here we examine additional limits and prospects with $8\tev$ and $13\tev$ searches in as model-independent a manner as possible.
~\\~\\~\noindent
{\bf $\gamma\gamma$ at $8\tev$:} 
Assuming 15 events are observed at $13\tev$ at $3.2\ifb$, we find to $\sigma_{\gamma\gamma}(13\tev)=4.6 \fb$. The gluon luminosity multiplier in going from 8 to $13\tev$ is 4.7 \cite{stirling}, giving $\sigma_{\gamma\gamma}(8\tev) \simeq 1\fb$. The increase in total luminosity from 13 to 8 affects both the signal and background, but the parton luminosity multiplier (appropriate for background) from 8 to 13 is order 2 as compared to 4.7 for the gluon luminosity from 13 to 8. We find that $\sigma_{\gamma\gamma}(8\tev) \simeq 1\fb$, which is not in conflict with the $8\tev$ search~\cite{Aad:2015mna}.
~\\~\\~\noindent
{\bf $Z\gamma$ at $8\tev$:}
Figure 3c of~\cite{Aad:2014fha} amusingly suggests a small bump at around 750GeV of about 0.35 fb for $S\rightarrow Z \gamma \rightarrow \ell\ell\gamma$.  Unfolding with respect to the branching ratio for $Z \rightarrow \ell\ell=0.06$ gives $S\rightarrow Z\gamma=5.75 \fb$.
For all the models we consider,
\begin{equation}
\frac{\sigma_{Z\gamma}}{\sigma_{\gamma\gamma}} =2t^2_W \simeq 0.6.
\end{equation}
Comparing to the estimate above, we expect $\sigma_{Z\gamma}(8\tev) \sim 0.6 \fb$, well below the $8\tev$ bounds indicated. Further, at $13\tev$ we expect $15\times0.6=9$ events at the current luminosity. However, the $Z\gamma$ channel is only efficiently probed via the $\ell\ell \gamma$ final state, providing a 6 percent branching ratio. Hence, for luminosity $\cL$, we expect the number of signal events to be $N_{Z\gamma}=15\times0.6\times0.06 \lesssim 1,$ which is not observable. 
~\\~\\~\noindent
{\bf Dijet bounds:}
The cross section times acceptance is bounded by $\sigma A \leq 1300 \fb$ from~\cite{Aad:2014aqa} for a scalar octet (we find that this provides stronger bounds than the updated search from \cite{ATLAS:2015nsi}). We conservatively assume similar bounds for a singlet particle decaying to two gluons.
\begin{equation}
\frac{\Gamma_{s \to gg}}{\Gamma_{s\to\gamma\gamma}}=x
\end{equation}
This translates to an $8\tev$ dijet cross section of $\sigma_{jj}= x\fb  \leq 1300 \fb $. This roughly corresponds to having $m_L \lesssim m_Q$ for $y_L \sim y_Q$. For fixed total width we dial both $y_L$ and $y_Q$ to the perturbative limit, the parameter space is symmetric in $m_Q$ and $m_L$ and hence this can be easily achieved.  In the ``QL Model'' we have seen that the viable parameter space is not strongly dependent on $m_Q$. Thus, $m_Q$ can be pushed up to suppress the dijet cross-section as needed.
~\\~\\~\noindent
{\bf Heavy Quarks:}
If the heavy quark $Q$ decays inside the instrument, for a vector-like quark with $q_Q=1/3, 2/3$, the bounds come from \cite{Khachatryan:2015oba} and \cite{Khachatryan:2015gza}, requiring $m_Q \gtrsim 750-920\gev$ depending on the channel. If instead the quarks are long lived, \cite{Khachatryan:2015jha,Aad:2013gva} rule out $m_Q \leq 500 \gev$. This is compatible with producing a sufficient number of diphoton events in our ``QL'' model.
~\\~\\~\noindent
{\bf Heavy Leptons:}
If the heavy lepton $L$ decays inside the instrument, the bounds depend on the decay channel. The bounds from \cite{Aad:2015dha} are somewhat weak, well below the requirement that $S$ not delay directly to $L$. If instead the leptons are long lived, \cite{Chatrchyan:2013oca,Aad:2015rba} rule out $m_L \leq 400 \gev$. Again, his is compatible with producing a sufficient number of diphoton events in our ``QL'' model.


\section{Conclusions}

We have demonstrated that a simple singlet scalar resonance with additional vector-like fermions charged under the Standard Model gauge group can account for the diphoton excess seen at ATLAS and CMS~\cite{atlasres,cmsres}. However, such a model shows tension with the large width preferred by ATLAS~\cite{atlasres}. If the width is not a fluctuation, this implies that the dominant branching fraction for this resonance is into states other that dijet and diphoton. Excitingly, we have further shown that the resonance must decay an $\mathcal{O}(1)$ fraction of the time into complicated visible sector states, as an invisible branching fraction that explains the width is ruled out. Additionally, to be compatible with the diphoton excess in a large width model requires both large couplings and vector like fermions with masses beneath the 750 GeV resonance. These fermions can be searched for directly depending on how the decays of the fermions are introduced in the model. However, these decay modes are not tied directly to the model for producing the diphoton excess, so searches for these consequences will be much more model dependent.

A model-independent way to obtain additional evidence for this model is in the $Z\gamma$ final state, which is always coupled to the number of observed photons (the dijet final is also interesting but not as clean and can be suppressed). Regardless of large or small width, we can rescale the ATLAS 8 TeV search for $Z\gamma$ resonances~\cite{Aad:2014fha} to predict the luminosity necessary for discovery. In this channel we expect $N_{\rm bgd} = 20 \times 2 \times \frac{3.2}{20} \times \frac{\cL}{3.2 \ifb}$. For a 3$\sigma$ discovery we find that we would need $\sim 600\ifb$ of data. Therefore in the first $300 \ifb$ we should expect hints in the $Z\gamma$ channel, and the high luminosity run of the LHC could definitively discover it in this channel. This would help to single out this explanation if the diphoton resonance persists, given the paucity of additional signals that a singlet scalar generates.

While more sophisticated explanations may describe the new diphoton excess, the model proposed in this letter is economical and generic. It points to interesting searches (e.g., in $Z\gamma$ and dijet final states) and highlights interesting tensions (e.g., with forcing a large branching fraction to invisible final states). Additional signals of new physics should be aggressively investigated in the context of the model-independent bounds advocated here and in more complete UV frameworks.

\section{Acknowledgements}

SDM thanks the Ronkonkoma branch of the Long Island Rail Road, where a portion of this work was completed. We thank R.~Essig, R.~Heller, D.~Tsybychev, and T.~Volansky for discussions, some of which were useful. The work of P.M. and H.R. was supported in part by NSF CAREER Award NSF-PHY-1056833. SDM is supported by NSF PHY1316617.

\end{document}